\documentclass[11pt]{article}
\usepackage{amsmath,amstext,amsbsy,amssymb}
\usepackage{bm}
\newcommand{\tr}{{\rm tr}}

\long\def\symbolfootnote[#1]#2{\begingroup%
\def\thefootnote{\fnsymbol{footnote}}\footnote[#1]{#2}\endgroup}

\begin{document}

\begin{center}

{\Large \bf   $3+1$ dimensional topological field theory as the effective action of neutral fermions }

\vspace{2cm}

\"{O}mer F. Dayi and Mahmut Elbistan

\vspace{5mm}

{\em {\it Physics Engineering Department, Faculty of Science and
Letters, Istanbul Technical University,\\
TR-34469, Maslak--Istanbul, Turkey}}\footnote{{\it E-mail addresses:} dayi@itu.edu.tr , elbistan@itu.edu.tr }

\end{center}

\vspace{2cm}

We proposed an action of neutral fermions  interacting with external electromagnetic fields to construct a $3+1$ dimensional
topological field theory  as the effective action attained  by
integrating out the fermionic fields in the related path integral.
These neutral quasiparticles are assumed to emerge from the collective behavior of the original physical particles and  holes
(antiparticles). Although our construction is general it is particularly useful to formulate effective actions of   the  time reversal invariant  topological insulators.

\vspace{2cm}

\section{Introduction}

Topological  field theories  by definition do not depend on the local features of the space-time in which they are defined.
They    appear in high energy physics \cite{bbrt} as well as in condensed matter physics. In the latter they arise as effective theories of quantum matter like topological insulators which 
 are usually defined to be ordinary insulators in the bulk but  conducting at the surface of the material (for a review see \cite{hk}) .
The material is under the influence of external electromagnetic fields.
In the low energy limit, 3+1 dimensional topological insulators subject to electromagnetic fields given by 
the gauge fields $A_\mu ;$ $\mu=0,1,2,3,$ are effectively described by
the action 
 $(\hbar =c=1),$
\begin{equation}
\label{seff4}
S_{3D}=\frac{\theta e^2}{8\pi^2}\int d^4x \ \epsilon^{\mu\nu\rho\sigma}\partial_\mu A_\nu \partial_\rho A_\sigma .
\end{equation}
It is topological in the sense that the metric tensor of the related space-time manifold does not appear in the action and there is no local excitations.
Properties of the underlying space-time should be fixed by additional information about the 
adopted microscopic model. In particle physics (\ref{seff4}) is known as the topological term for the $\theta$-vacuum \cite{cdg,jr}.
For  compact space-time manifolds by choosing $\theta =\pm (2n+1)\pi ;\ n=0,1,\cdots ,$ (\ref{seff4}) describes
the main features of the time reversal invariant  topological insulators in $3+1$ dimensions \cite{qhz}.

One of the models which gives rise to 
the  action (\ref{seff4})
is obtained through a dimensional reduction from the $4+1$ dimensional Dirac theory \cite{qhz}.
One considers  the relativistic electrons  
interacting with the external gauge fields $A_\mu,$ as well as with  the scalar field 
$\Theta (x) $  described by 
\begin{equation}
\label{idm}
{\cal L}_{3+1}(\Psi,\bar{\Psi},A)=\bar{\Psi}[i\gamma^{\mu}(\partial_\mu-i eA_\mu)+ \gamma^4(k_4+\Theta )-m]\Psi .
\end{equation}  
Here, $k_4$ is a constant and we consider electrons of charge $e>0.$   
$\gamma_4$ is one of the  gamma matrices introduced in $4+1$ dimensions and
$\Theta(x)$ is the reminiscent of the gauge field $A_4(x).$ 
Integrating out the fermionic fields $\Psi,\bar{\Psi} $ in the path integral  of (\ref{idm}),
yields the effective action \cite{qhz,dey}
\begin{equation}
\label{aa}
S_A=
\frac{ e^2}{8\pi^2}\int d^4x \ \Theta (x) \epsilon^{\mu\nu\rho\sigma}\partial_\mu A_\nu \partial_\rho A_\sigma .
\end{equation}
$\Theta (x)$   is known in particle physics as axion field (for a review see \cite{kim}). For a uniform and constant axion field 
$\Theta = \theta ,$ (\ref{aa})  leads to  (\ref{seff4}). 

The topological field theory action (\ref{seff4}) is also related to the phase of  mass term of quarks  through the axial anomaly \cite{baluni} (for a recent discussion see \cite{hjk}). One deals with the fermionic fields with negative mass described by the Lagrangian density
$$
\label{nma}
{\cal L}_{MD}= \bar{\psi}_M\left[i\gamma^{\mu}(\partial_{\mu}-i eA_\mu) -\exp (-i\pi)M\right]\psi_M.
$$ 
In terms of axial transformations one can obtain a positive mass term. Thus,
integrating out the $\bar{\psi}_M, \psi_M$ fields in the path integral, with an appropriate  renormalization procedure,
one  obtains the action (\ref{seff4}) for $\theta=\pi.$ 

We would like to present another strictly $3+1$ dimensional  model which leads to (\ref{seff4}). We will introduce an action of neutral  fermionic quasiparticles and show that by integrating out these fermionic fields one can  obtain  (\ref{seff4})  through  a renormalization procedure. 
These neutral fermions are not the fundamental particles but  arise as excitations made of particle and hole (antiparticle)  solutions of the underlying Dirac theory of electrons.   It will be shown that our  approach can also be generalized to
derive the $BF$ type theory which was proposed to describe topological insulators in \cite{cm}.

How to construct these neutral fermions from the underlying charged  particles is not known. However, 
neutral fermion excitations have already been appeared in the study of the Hall effect for even filling fractions  \cite{mr, gww}. Moreover, 
neutral Dirac fermions composed of two  Majorana fermions arise in systems composed of topological, magnetic and superconducting insulators \cite{fk}. 

\section{Effective field theory 
\label{s2}}

Let us deal with the electrons denoted   $\psi_e.$ They interact with  external electromagnetic fields  according to the Dirac Lagrangian  density
\begin{equation}
\label{DL}
{\cal L}_D=  \bar{\psi}_e\left[i\gamma^{\mu}(\partial_{\mu}-i eA_\mu) -m\right]\psi_e.
\end{equation}
On-shell states satisfy the Dirac equation. As is well known, there are negative energy solutions of  the Dirac equation 
describing positive charged particles which can be interpreted as antiparticles (holes). Hence, when we deal with 
the excitations of the Dirac particles we may consider particles of both sign. They 
can be tight together to form neutral degrees of freedom. In particular we may  assume that
in the bulk of a topological insulator effectively there are neutral fermions. This guarantees that there is no electrical transport
in the bulk, which is the basic property of an insulator.
Obviously, neutral fermions 
couple to  external electromagnetic fields due to their electric and magnetic dipole moments whose interaction terms covariantly given by $ \sigma^{\mu\nu}F_{\mu\nu}$ and $i\gamma^5\sigma^{\mu\nu}F_{\mu\nu},$ where 
$$ 
F_{\mu\nu}=\partial_{\mu}A_{\nu}-\partial_{\nu}A_{\mu};\ \ \ 
\sigma^{\mu\nu}=\frac{i}{2}[\gamma^\mu,\gamma^\nu];\ \ \ \gamma_5=i\gamma^0\gamma^1\gamma^2\gamma^3.
$$
We propose that  the neutral fermions are described by the Lagrangian density
\begin{equation}
\label{Ladi4}
{\cal L}(\bar{\psi}, \psi, F)=\bar{\psi}\left[i\gamma^{\mu}\partial_{\mu}+\frac{1}{2}e\alpha(1-i\gamma^5)\sigma^{\mu\nu}F_{\mu\nu}-m\right]\psi,
\end{equation} 
where  $\alpha$ is a constant. In the related path integral one can integrate out the fermions
to derive 
the effective action of the electromagnetic field strengths $S[F]$: 
$$
\exp(iS[F])\equiv \int d\bar{\psi} d\psi \exp\left(i\int d^4x {\cal L}(\bar{\psi}, \psi, F)\right).
$$
 Formally, the effective action can be written as
\begin{equation}
\label{det}
S[F]=-i\ln\det[i\gamma^{\mu}\partial_{\mu}+\frac{1}{2}e\alpha(1-i\gamma^5)\sigma^{\mu\nu}F_{\mu\nu}-m].
\end{equation}
As usual, it is preferable to work in the momentum space.
One of the terms which (\ref{det}) generates  is
\begin{equation}
\label{fre}
\int \ \frac{d^4k_1}{(2\pi)^4}\frac{d^4k_2}{(2\pi)^4} F_{\mu\nu}(k_1)F_{\rho\sigma}(k_2)\pi^{\mu\nu\rho\sigma}(k_1)\delta^4(k_1-k_2).
\end{equation}
We deal with the low energy limit where momentum of the external legs of the related Feynman diagram vanishes. In this limit
(\ref{fre}) can be taken as the effective action. 
At the one loop level  $\pi^{\mu\nu\rho\sigma}(k)$ can be written as
\begin{equation}
\label{pi} 
\pi^{\mu\nu\rho\sigma}(k)=\frac{i}{2}e^2\int\frac{d^4p}{(2\pi)^4}\tr[(\frac{\alpha}{2}(1-i\gamma^5)\sigma^{\mu\nu}S_F(p))(\frac{\alpha}{2}(1-i\gamma^5)\sigma^{\rho\sigma}S_F(p-k))],
\end{equation}
where the Feynman propagator is defined as 
$$
S_F(p)=\frac{i(\gamma^{\mu}p_{\mu}+m)}{p^2-m^2} .
$$
Using the  properties of Dirac matrices in $3+1$ dimensions, one can show that (\ref{pi}) reduces to
two terms:
\begin{equation}
\label{tpi}
\pi^{\mu\nu\rho\sigma}(k)=\Sigma^{\mu\nu\rho\sigma}(k)+\Pi^{\mu\nu\rho\sigma}(k),
\end{equation}
which are defined  as 
\begin{eqnarray*}
\Sigma^{\mu\nu\rho\sigma}(k) & = & \frac{i}{16}e^2\alpha^2\int \frac{d^4p}{(2\pi)^4}\frac{p_{\alpha}(p-k)_{\beta}\tr\{[\gamma^{\mu},\gamma^{\nu}]\gamma^{\alpha}[\gamma^{\rho},\gamma^{\sigma}]\gamma^{\beta} }{(p^2-m^2)((p-k)^2-m^2)},\\ 
\Pi^{\mu\nu\rho\sigma}(k) & = & \frac{1}{16}e^2\alpha^2m^2\int \frac{d^4p}{(2\pi)^4}\frac{
\tr\{[\gamma^{\mu},\gamma^{\nu}][\gamma^{\rho},\gamma^{\sigma}]\gamma^5 \}}{(p^2-m^2)((p-k)^2-m^2)} .
\end{eqnarray*} 
$\Sigma^{\mu\nu\rho\sigma}(k)$  is  quadratically divergent. 
In the low energy limit $(k\rightarrow 0)$ it leads to 
$$
\int \frac{d^4p}{(2\pi)^4}\frac{p_\alpha p_\beta}{(p^2-m^2)^2}=\left[\frac{\Lambda^2}{16 \pi^2} +\frac{m^2}{4\pi^2} 
\ln \left(\frac{\Lambda}{m}\right) +\cdots\right]
\eta_{\alpha\beta},
$$ 
where $\eta_{\alpha\beta}$ is the  metric tensor and  $\Lambda$ is the ultraviolet cut-off. Although this term is divergent for 
$\Lambda \rightarrow \infty,$ it is multiplied with
the  trace term which can be seen to vanish in 4 dimensions:
$$
\tr\{[\gamma^\mu,\gamma^\nu]\gamma^\alpha[\gamma^\rho,\gamma^\sigma]\gamma_\alpha\}=0.
$$
Thus we set
$$
\label{sigma}
\Sigma^{\mu\nu\rho\sigma}(0)= 0.
$$ 

The other term in (\ref{tpi})
is topological, i.e. it is independent of the space-time metric $\eta_{\mu\nu}. $ In terms of the totally antisymmetric tensor $\epsilon^{\mu\nu\rho\sigma}$ originated from 
$$
\tr\left\{\gamma^\mu\gamma^\nu\gamma^\rho\gamma^\sigma\gamma^5\right\}=-4i\epsilon^{\mu\nu\rho\sigma},
$$
it can be expressed as
$$
\Pi^{\mu\nu\rho\sigma}(k)=e^2m^2\alpha^2\epsilon^{\mu\nu\rho\sigma}I(k).
$$
We defined
\begin{equation}
\label {Ik}
I(k)=-i\int\frac{d^4p}{(2\pi)^4}\frac{1}{((p^2-m^2)((p-k)^2-m^2)},
\end{equation}
whose  integrand is spherically symmetric
in the low energy limit $(k\rightarrow 0).$  
The integral is logarithmically divergent which can be  dimensionally regularized. To calculate (\ref{Ik}) in $d$ dimensions we introduce the parameter  $\mu$ having the dimension of  mass such that     the integral
\begin{equation}
\label{iz}
I(0)=-i\mu^{4-d}\int\frac{d^dp}{(2\pi)^4}\frac{1}{(p^2-m^2)^2},
\end{equation}
remains dimensionless.
Calculation of $I(0)$ in $d$ dimensions leads to \cite{ps}
$$
I(0)= \frac{1}{4\pi^{d/2}}\Gamma(2-d/2) (\frac{\mu^2}{m^2})^{2-d/2},
$$
where $\Gamma(x)$ is the gamma function.  

Setting $d=4-\varepsilon$  the  finite and infinite parts in the $\epsilon \rightarrow 0$ limit, are separated as
$$
I(0)=\frac{1}{(4\pi)^2}(\frac{2}{\epsilon}+\ln(\frac{\mu^2}{m^2})+\ln(4\pi)-\gamma+...).
$$
$\gamma$ is the Euler-Mascheroni constant. 
The bare coupling constant  $\alpha,$ or the mass $m,$ 
can be renormalized  with the divergent constant $Z$ to introduce  the finite parameter $ \theta_F$ as
$$
Zm^2\alpha^2I(0)=\frac{\theta_F}{32\pi^2}.
$$ 
As it is usual
in the renormalization theory,
the value of  $ \theta_F$  will be fixed in terms of the ``experimentally measured quantities."  We conclude that
$$
\pi^{\mu\nu\rho\sigma}(0)=\frac{e^2\theta_F}{32\pi^2} \epsilon^{\mu\nu\rho\sigma}.
$$
Therefore, in the low energy limit the renormalized effective action is
\begin{equation}
\label{rac}
S[F]=\frac{e^2\theta_F}{32\pi^2} \int d^4x \epsilon^{\mu\nu\rho\sigma}F_{\mu\nu}(x)F_{\rho\sigma}(x).
\end{equation}
It is the same  with the  action (\ref{seff4}) for $\theta_F=\theta.$

Properties of the space-time on which the theory is defined should be dictated by the physical arguments.
We assume that the neutral quasiparticle of the initial Lagrangian density (\ref{Ladi4}) is composed of the original  particles and holes. 
For example in the case of topological insulators
the physical particles  are the electrons
minimally  coupled to the gauge fields $A_\mu $ as  in (\ref{DL}). Then, there are some configurations which are consistent with periodic space-time yielding the quantization condition
\begin{equation}
\label{q1}
\frac{e^2}{32\pi^2}\int d^4x \ \epsilon^{\mu\nu\rho\sigma}F_{\mu\nu}F_{\rho\sigma} = N,
\end{equation}
where $N$ is an integer. Thus the partition functions for $\theta_F =\pm (2n+1)\pi ;\ n=0,1,\cdots , $ are the same, 
so that the theory is time reversal invariant. 

Obviously, there is no a priori given condition for the space-time structure of the topological field theory action (\ref{rac}).
We may choose  the space-time to be  periodic but yielding the quantization condition
\begin{equation}
\label{q2}
\frac{e^2}{32\pi^2}\int d^4x \ \epsilon^{\mu\nu\rho\sigma}F_{\mu\nu}F_{\rho\sigma} = N_f^2N,
\end{equation}
where $N_f$ is an odd integer.
Then, we set  $\theta_F=\pm (2n+1)\pi/N_f;\ n=0,1,\cdots ,$ which defines  
the time reversal invariant fractional topological insulator \cite{mqkz,sbms,ph}. In this case, 
as in \cite{mqkz}, the partons denoted  $\psi_p,$ 
 whose electromagnetic  interaction Lagrangian density is given by
$$
\label{PL}
{\cal L}_p=  \bar{\psi}_p\left[i\gamma^{\mu}(\partial_{\mu}-i \frac{e}{N_f} A_\mu ) -m\right]\psi_p ,
$$
can be chosen
as the physical  particles in the bulk of the fractional topological insulator. They possess the electric charge $e/N_f,$
so that  the quantization condition (\ref{q2}) is justified.

\section{$BF$ theory }

Another interesting topological field theory is the BF theory \cite{bbrt}. It also appears  in condensed matter physics. 
In addition to the electromagnetic gauge field $A_{\mu},$ let us introduce the auxiliary fields  $a_{\mu},\ b_{\rho\sigma}$ which are Abelian vector and  antisymmetric tensor fields.  It was argued in \cite{cm}
that the time reversal invariant $3+1$ dimensional topological insulator is  described by the $BF$ type  effective field theory given by  
\begin{equation}
\label{LBF}
{\cal L}_{BF}=\frac{e}{2\pi}\epsilon^{\mu\nu\rho\sigma}a_{\mu}\partial_{\nu}b_{\rho\sigma}+\frac{e}{2\pi}\epsilon^{\mu\nu\rho\sigma}A_{\mu}\partial_{\nu}b_{\rho\sigma}+C \epsilon^{\mu\nu\rho\sigma}\partial_{\mu}a_{\nu}\partial_{\rho}A_{\sigma},
\end{equation}
where $C$ is a constant parameter. To fix  the value of it as $C=\pm e^2/ 8\pi ,$ the gauge charge lattice was  specified as an additional information. Thus, integrating out  $a_{\mu}$ and $b_{\rho\sigma}$ fields in the related path integral  yields 
\begin{equation}
\label{pm3d}
S_{3D}=\pm \frac{e^2}{8\pi}\int d^4x\epsilon^{\mu\nu\rho\sigma}\partial_{\mu}A_{\nu}\partial_{\rho}A_{\sigma},
\end{equation}
which is the same with (\ref{seff4}) for $\theta=\pm 1.$

It is possible to construct (\ref{LBF}) in terms of the procedure outlined in  Section \ref{s2}. For this purpose let us  introduce the neutral fermions denoted $\psi_A;\ A=1,2,\cdots ,6$ and
define the Lagrangian density 
\begin{eqnarray*}
{\cal L}&=&
\bar{\psi}_1\left[i\gamma^{\mu}\partial_{\mu} +\frac{1}{2}\lambda(1-i\gamma^5)\sigma^{\mu\nu}(F_{\mu\nu}+b_{\mu\nu})-m\right]\psi_1 \\
&+&\bar{\psi}_2\left[i\gamma^{\mu}\partial_{\mu} +\frac{1}{2}\lambda(1+i\gamma^5)\sigma^{\mu\nu}(F_{\mu\nu}-b_{\mu\nu})-m\right]\psi_2 \nonumber \\ 
&+&\bar{\psi}_3\left[i\gamma^{\mu}\partial_{\mu} +\frac{1}{4}\beta(1-i\gamma^5)\sigma^{\mu\nu}(F_{\mu\nu}+f_{\mu\nu})-m\right]\psi_3\\
&+&\bar{\psi}_4\left[i\gamma^{\mu}\partial_{\mu} +\frac{1}{4}\beta(1+i\gamma^5)\sigma^{\mu\nu}(F_{\mu\nu}-f_{\mu\nu})-m\right]\psi_4\nonumber \\
&+&\bar{\psi}_5\left[i\gamma^{\mu}\partial_{\mu} +\frac{1}{2}\lambda(1-i\gamma^5)\sigma^{\mu\nu}(f_{\mu\nu}+b_{\mu\nu})-m\right]\psi_5\\
&+&\bar{\psi}_6\left[i\gamma^{\mu}\partial_{\mu} +\frac{1}{2}\lambda(1+i\gamma^5)\sigma^{\mu\nu}(f_{\mu\nu}-b_{\mu\nu})-m\right]\psi_6,
\end{eqnarray*}
where  $f_{\mu\nu}=\partial_{\mu}a_{\nu}-\partial_{\nu}a_{\mu}.$ 
Now, we can integrate out each $\psi_A, \bar{\psi}_A$ neutral fermion  degrees of freedom 
following the procedure of Section \ref{s2}. Obviously, at the one loop level integrals are divergent
which can be expressed in the low energy limit as $I(0)$ which is defined in (\ref{iz}). 
The bare coupling constants $\lambda$ and $\beta $ can be  renormalized
to define the finite constants   $\Lambda_F$ and $C_F$ as
$$
Zm^2\lambda^2I(0)=\Lambda_F,\ \ Zm^2\beta^2I(0)=C_F.
$$
Then, we can write the low energy effective action as
\begin{equation}
\label{rnca}
S[A,b,a]=\int \ d^4x \left\{\Lambda_F\left[ \epsilon^{\mu\nu\rho\sigma}\partial_{\mu}A_{\nu}b_{\rho\sigma}+\epsilon^{\mu\nu\rho\sigma}\partial_{\mu}a_{\nu}b_{\rho\sigma}\right]+C_F\epsilon^{\mu\nu\rho\sigma}\partial_{\mu}A_{\nu}\partial_{\rho}a_{\sigma}\right\}.
\end{equation}
As usual the renormalized quantities  should be fixed by some additional conditions. 
We can choose  $\Lambda_F=1/2\pi$ and $C_F=\pm e^2/8 \pi ,$ so that (\ref{rnca}) yields the $BF$ theory described  by  (\ref{LBF}), up to surface terms. Obviously,
integrating out first $b_{\mu\nu}$ then $a_\mu$ fields leads to  the action (\ref{pm3d}). 

\section{Discussions}

We constructed in the low energy limit the effective action  (\ref{rac}),
by integrating out the neutral Dirac particles described by the Lagrangian density (\ref{Ladi4}). 
We assume that  the neutral fermions are formed as excitations of the
original physical particles which are electrons or partons.
Our procedure of calculating  the effective topological field theory (\ref{rac})
does not refer to the origin of the neutral fermions. 
The original physical particles establish the features of the space-time manifold on which the effective topological field theory is defined. Then, according to the quantization conditions like (\ref{q1}) and (\ref{q2}),  
the value of the physical coupling constant $\theta_F$ should be chosen.

Interchanging Abelian field strengths with non-Abelian ones in the   Lagrangian density (\ref{Ladi4}) will not alter the construction of the low energy effective action (\ref{rac}) presented in Section \ref{s2}. Therefore, our procedure is also valid for non-Abelian gauge fields. In this case, quantization condition of the effective action will be  changed  \cite{witten}, so that the renormalized coefficient $\theta_F$ should be appropriately  chosen.

Although, mechanism of constructing neutral fermions from the original ones is 
an open issue, our procedure indicates that different kind of topological insulators  can be studied in a unified
manner.

\newpage

\newcommand{\PRL}{Phys. Rev. Lett. }
\newcommand{\PRB}{Phys. Rev. B }

\end{document}